# A Step Forward from High-Entropy Ceramics to Compositionally Complex Ceramics: A New Perspective

Andrew J. Wright[a], Jian Luo[a,b,*]

[a]Department of NanoEngineering; [b]Program of Materials Science and Engineering, University of California, San Diego, La Jolla, CA 92093, USA

## Abstract

High-entropy ceramics (HECs) have quickly gained attention since 2015. To date, nearly all work has focused on five-component, equimolar compositions. This perspective article briefly reviews different families of HECs and selected properties. Following a couple of our most recent studies, we propose a step forward to expand HECs to Compositionally Complex ceramics (CCCs) to include medium-entropy and non-equimolar compositions. Using defective fluorite and ordered pyrochlore oxides as two primary examples, we further consider the complexities of aliovalent cations and anion vacancies as well as ordered structures with two cation sublattices. Better thermally-insulating yet stiff CCCs have been found in non-equimolar compositions with optimal amounts of oxygen vacancies and in ordered pyrochlores with substantial size disorder. It is demonstrated that medium-entropy ceramics (MECs) can prevail over their high-entropy counterparts. The diversifying classes of CCCs provide even more possibilities than HECs to tailor the composition, defects, disorder/order, and, consequently, various properties.

[*] Correspondence should be addressed to J. Luo (jluo@alum.mit.edu).

# Graphical Abstract

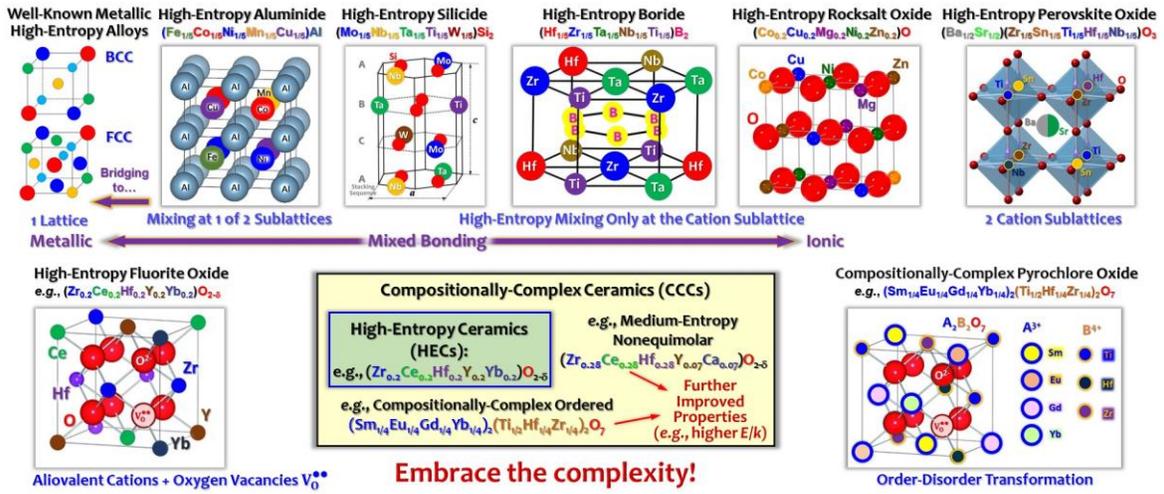

# 1. Introduction

High-entropy ceramics (HECs), while still in their nascent beginning, are developing as the ceramic counterparts to the more mature high-entropy alloys (HEAs) [1–4]. Despite some earlier related reports, the field of HEAs emerged in 2004 with the seminal publications of Yeh et al. [5] and Cantor et al. [6] demonstrating the feasibility of alloying five or more elements at equimolar fractions. Albeit some earlier studies on nitride and other thin films [5, 7–10], the HECs have simulated considerable interests in the ceramics community since the report of a bulk entropy-stabilized oxide (ESO) $(Co_{0.2}Cu_{0.2}Mg_{0.2}Ni_{0.2}Zn_{0.2})O$ by Rost et al. in 2015 [11]. Since then, work has grown exponentially with numerous efforts of synthesizing different families of HECs (Fig. 1) and exploring their properties. Zhang and Reece [1] have recently reviewed the design, synthesis, structure, and properties of HECs. After the initial submission of this article, Oses et al. [12] also published a review of HECs. Thus, this article will not repeat a comprehensive review to avoid redundancy; instead, our focus is on discussing a new perspective that emerged by assembling and synthesizing new ideas from the discoveries made in a couple of very recent research reports [13, 14], as discussed below.

Early work on HEAs began as equimolar compositions [3]. More recently, studies on HEAs have reported the benefits of medium-entropy and non-equimolar compositions in improving mechanical properties [15–17]. The development of HECs is experiencing a similar pathway as most studies to date have been focused on five-component (and occasionally four-component) equimolar compositions. In 2020, Wright et al. first proposed to expand HECs to Compositionally Complex ceramics (CCCs) to include medium-entropy and/or non-equimolar compositions [13, 14]. In one example of yttria-stabilized zirconia (YSZ)-like defective fluorite oxides, non-equimolar medium-entropy compositions were found to exhibit further reduced thermal conductivity in comparison with their high-entropy equimolar counterparts, presumably due to the effects of oxygen vacancies clustering [13]. In another case of ordered pyrochlore oxides, reduced thermal conductivity was found to be correlated better with size disorder, instead of ideal mixing entropy itself; thus, medium-entropy compositions can again outperform their high-entropy counterparts [14].

In this short perspective article, we first briefly review and discuss the recent discoveries of different families of HECs. Subsequently, an in-depth analysis of thermally-insulative yet hard and stiff HECs/CCCs is given as an example to highlight the new opportunities. Following our two recent reports [13, 14], a particular goal is to elaborate further the proposal of broadening HECs to CCCs as a step forward. We also discuss the diversifying classes of CCCs that provide even more possibilities than HECs to tailor the composition, defects, disorder, and order to achieve better and more tunable properties.

**2. Terminologies and Classifications: A Step Forward from HECs to CCCs**

Several definitions of metallic HEAs have been summarized and discussed by Miracle and Senkov [3]. Similar to their metallic counterparts, we may (loosely) define High-Entropy Ceramics or HECs as compositions of five or more principal (~5%–35%) cations (often in equimolar or near-equimolar fractions), with ideal mixing entropy of greater than $1.5k_B$ per cation (on at least one cation sublattice if there are two or more cation sublattices), where $k_B$ is the Boltzmann constant.

A phase is considered to be "entropy-stabilized" when the entropy contributions overcome an enthalpic barrier, as demonstrated by Rost et al. [11]. Correspondingly, a class of entropy-stabilized ceramics (ESCs, with the ESOs as a subclass [11]) can be defined. We should note that the concept of "entropy-stabilized phase" and ESCs can be applied to cases beyond the high-entropy compositions. Discussion of ESCs is not the focus of this perspective article.

In a most recent study of fluorite oxides, Wright et al. first proposed to extend HECs to Compositionally Complex Ceramics or CCCs to include medium-entropy and non-equimolar compositions (Fig. 2) [13]. This is similar to the terminologies of Compositionally Complex or complex-concentrated alloys (CCAs) used in the physical metallurgy community. CCCs may alternatively be named as "multi-principal cation ceramics (MPCCs)," analogous to their metallic counterpart, multi-principal element alloys (MPEAs).

Here, CCCs include medium-entropy ceramics (MECs) that typically have mixing configurational entropy in the range of $1 - 1.5k_B$ per cation (on at least one cation sublattice if there are multiple cation sublattices), with similar definitions used for their metallic counterparts [18–22]. MECs with one cation sublattice include (*i*) 3-4 cation equimolar (or near-equimolar) compositions, e.g., $(Hf_{1/4}Zr_{1/4}Ce_{1/4}Y_{1/4})O_{2-\delta}$ [23] and $(Ta_{1/3}Zr_{1/3}Nb_{1/3})C$ [22] and (*ii*) non-equimolar compositions with 3-4 principal plus a few minor (typically <5%) cations, e.g., $(Hf_{0.314}Zr_{0.314}Ce_{0.314}Y_{0.029}Ca_{0.029})O_{2-\delta}$ [13] (Fig. 2). Also, ordered MECs with two sublattices include $Gd_2(Sn_{1/4}Ti_{1/4}Hf_{1/4}Zr_{1/4})_2O_7$ and $(Sm_{1/2}Gd_{1/2})_2(Ti_{1/3}Hf_{1/3}Zr_{1/3})_2O_7$ [14].

We should note that we use "ideal" mixing entropy here to define high- vs. medium-entropy compositions, as some chemical (cation) short-range orders (CSROs) like exist and reduce the actual mixing (configurational) entropy.

Additional complexity arises for CCCs with two (or more) cation sublattices, e.g., perovskite ($ABO_3$) [24–28], spinel ($AB_2O_4$) [29–32], and pyrochlore ($A_2B_2O_7$) [33–37] oxides. Here, we typically distinguish HECs and MECs based on high- or medium-entropy mixing on one of the cation sublattices (typically according to the one with the highest ideal mixing entropy). For HECs with two cation sublattices, high-entropy mixing can occur at one cation sublattice, e.g., in $(La_{1/5}Ce_{1/5}Nd_{1/5}Sm_{1/5}Eu_{1/5})_2Zr_2O_7$ [14, 34] and $(Ba_{1/2}Sr_{1/5})(Zr_{1/5}Sn_{1/5}Ti_{1/5}Hf_{1/5}Nb_{1/5})O_3$ [26], or on both cationic sublattices, e.g., in $(Gd_{1/5}La_{1/5}Nd_{1/5}Sm_{1/5}Y_{1/5})(Co_{1/5}Cr_{1/5}Fe_{1/5}Mn_{1/5}Ni_{1/5})O_3$ [25] (Fig. 2).

A further somewhat vague case is represented in CCCs with medium-entropy mixing at two cation sublattices, *e.g.*, in $(Sm_{1/4}Eu_{1/4}Gd_{1/4}Yb_{1/4})_2(Ti_{1/4}Sn_{1/4}Hf_{1/4}Zr_{1/4})_2O_7$ and $(Sm_{1/3}Eu_{1/3}Gd_{1/3})_2(Ti_{1/2}Sn_{1/6}Hf_{1/6}Zr_{1/6})_2O_7$ [14], so that they can have an overall mixing entropy higher than some of those having high-entropy mixing on only one of two cation sublattices but low mixing entropy on the other cation sublattice; yet these CCCs still have $< 1.5 k_B$ per cation on average (or on any cation sublattice). Thus, they may be classified as HECs or MECs (somewhat subjectively).

We should note the boundaries of and between HECs, MECs, and CCCs are rather vague (not rigorously defined), and definitions are mostly subjective.

Further extensions include mixing-anion CCCs [38–42] and possibly covalent (and, very often, mixing bonding) CCCs.

## 3. Overview of HECs

### 3.1 Systems

Recently, Zhang and Reece comprehensively reviewed the HEC systems and synthesis methods [1]. Here, we briefly discuss selected essential progress. Fig. 1 illustrates selected high-entropy materials with different bonding nature and crystal structures, from metallic HEAs, which are mostly in simple BCC and FCC, and some HCP structures [3], and high-entropy aluminides with mostly metallic (but some ionic) bonding and ordered two sublattices [43], to mixed (metallic/covalent/ionic) bonding in high-entropy silicides [44, 45] and borides [46–56] (as well as carbides [21, 22, 57–66] and nitrides [67] not shown here), to ionic high-entropy oxides [11, 13, 14, 23–37, 68–96].

Within the high-entropy oxides, most initial attention [68–74, 77–81, 83, 85, 86, 88–93, 95] has focused on the rocksalt ESO $(Mg_{1/5}Ni_{1/5}Co_{1/5}Cu_{1/5}Zn_{1/5})O$ first reported by Rost et al. [11]. Other oxide systems of significant interests include those with the fluorite [13, 23, 75, 76, 84], spinel [29–32, 96], pyrochlore [14, 33–37], and perovskite [24–28] structures.

Pyrochlore and perovskite oxides are unique because medium- or high-entropy mixing can be achieved on multiple cation sites (Fig. 1) [14, 24–28, 33–37]. Additionally, both fluorite/pyrochlore and rocksalt/spinel oxides can undergo order-disorder transformations, which have not yet been explored for medium- and high-entropy compositions; nevertheless, it offers potentially a new avenue to further engineer CCCs.

In 2016, Gild et al. [48] first reported the fabrication of high-entropy borides (metal diborides of the $AlB_2$ structure) as a new class of high-entropy ultra-high temperature ceramics (UHTCs). These high-entropy borides are interesting because they have mixed covalent, metallic, and ionic bonds, with a unit layered hexagonal crystal structure consisting of a 2D high-entropy mixing of metal/cation atoms separated by rigid covalently-bonded boron nets (Fig. 1). Although the initial work on high-entropy borides resulted in relatively low densities (~ 92%) due to significant oxide contamination from high-entropy ball milling, several subsequent studies quickly improved powder synthesis and fabrication methods that enhanced relative densities and properties [46, 47, 49–56]. In 2018 and 2019, several groups [21, 22, 57–66] also independently reported the fabrications of high-entropy carbides as another subclass of high-entropy UHTCs. High-entropy boride-carbide two-phase UHTCs have also been fabricated and examined recently [97]. However, the bulk mechanical properties such as flexural strength and fracture toughness (other than indentation toughness) need to be tested to further develop and enable these high-entropy UHTCs for applications in extreme environments.

In addition to two major classes high-entropy UHTCs (discussed above) that have been extensively studied in the last a few years, high-entropy nitrides [67], silicides [44, 45], sulfides [98], fluorides [99], aluminides [43], hexaborides [100], carbonitrides [101], and alumino-silicides [38] have been fabricated. In the broader families of oxide-related HECs, the fabrication of high-entropy magnetoplumbites [87, 102], zeolitic imidazolate frameworks [103], ferrites [104], phosphates [18, 105], monosilicates [19, 20], disilicates [106], and metal oxide nanotube arrays [107] have been reported. Medium- and high-entropy Compositionally Complex thermoelectrics have also been explored [40–42]. Most of these studies found homogeneously distributed cations demonstrating the formations of high-entropy solid solutions.

## 3.2 Modeling

Modeling is critical to help further the understanding of HECs. Notably, Sarker et al. established a descriptor to help predict single-phase formation in high-entropy carbides from density-functional theory calculations [59, 66]. Efforts are being made to extend and validate the descriptor for a broad range of other material systems. Additionally, modeling has revealed the importance of size and interatomic force constant disorder resulting in thermally-insulative HECs [78, 108–110]. Various other modeling studies have been conducted [60, 111–115]. A complete review and critical assessment of the modeling of HECs is beyond the scope of this perspective article. In a review published after the initial submission of this article, Oses et al. [12] provided a brief summary of modeling studies (along with other aspects of HECs).

## 3.3 Properties

Among the oxides, some interesting and intriguing functional properties discovered so far include: high dielectric constants [70] and lithium-ion conductivity [69, 83, 93, 116], low-temperature water splitting [82], stable, high-temperature catalytic properties [79], pressure-induced amorphization [117], and tunable magnetism [28, 31, 32, 85, 88]. The properties of HECs have been reviewed previously [1, 2, 12, 116].

The high-entropy borides and carbides are being examined for their potential use as next-generation UHTCs [118, 119]. These classes of materials have also shown increased mechanical properties [56, 61, 62, 64, 66] and oxidation resistance [48, 63, 120, 121] compared to their constituents or a rule of mixtures (RoM) analysis.

Notably, a general property of HECs is represented by the increased hardness in comparison with the RoM averages, which have been reported for high-entropy borides [48, 50, 54, 56], carbides [59, 61, 63, 66], and silicides [44, 45]. Further discussion can be found in the next section. Without a surprise, HECs also generally exhibit reduced thermal conductivity due to the increased phonon scattering, which will be discussed in more detail in the next section.

Another possible general property of HECs and CCCs is the increased phase stability for the high-symmetry phase. Specifically, the phase transformation temperature from a high-symmetry, high-$T$ (entropy-stabilized) phase to a low-symmetry, low-$T$ (enthalpy-stabilized) phase may be reduced. In other words, the high-$T$, high-symmetry phase may be stabilized to lower temperatures (for a large temperature region) in HECs/CCCs in comparison with their low-entropy counterparts. Two recent examples are given below. For example, Liu et al. found the monoclinic $Cu_2$(S/Se/Te) system transformed into a hexagonal structure at room temperature when the configurational entropy in was greater than ~ $0.5R$ per mol. [40]. A similar effect has also been reported by Wright et al. in YSZ-like Compositionally Complex fluorite oxides [13]. While 3YSZ, $(Zr_{0.942}Y_{0.058})O_{2-\delta}$, undergoes a tetragonal to cubic transition around 2200°C, its high-entropy counterpart $(Hf_{0.314}Ce_{0.314}Zr_{0.314}Y_{0.029}Yb_{0.029})O_{2-\delta}$, which has an identical concentration of cubic stabilizers, experienced this transition around 1400°C, representing a remarkable reduction of the phase transformation temperature of ~800°C (*i.e.*, an increase of the stability region of the high-$T$ cubic phase) [13].

## 4. An Example of New Opportunities: Thermally-Insulating Yet Stiff CCCs

Persistent in a broad range of HECs are enhanced mechanical properties [20, 22, 44, 48, 50, 54, 56, 59, 61–64, 66, 78] and diminished thermal conductivity [20, 23, 33, 34, 36, 40–42, 44, 54, 65, 78, 84, 98, 105, 122], in particular, with respect to their RoM averages. This enables HECs to possess a new unique property since Young's modulus and thermal conductivity are typically inversely correlated, as shown in Fig. 3 (replotted after Braun et al. [78]). In Fig. 3, we have further added our data points for three Compositionally Complex fluorite-based oxides: (1) "YSZ-like" non-equimolar $Hf_{0.284}Zr_{0.284}Ce_{0.284}Y_{0.074}Yb_{0.074}O_{2-\delta}$ [13] and (2) $(YDyErNb_{1/2}Ta_{1/2})O_7$ rare earth niobate/tantalate [Wright, Wang, Chen and Luo, unpublished results], both of which are in the disordered fluorite structure, as well as (3) $(Sm_{1/4}Eu_{1/4}Gd_{1/4}Yb_{1/4})_2(Ti_{1/2}Hf_{1/4}Zr_{1/4})_2O_7$ in the ordered pyrochlore structure [14], in addition to rocksalt ESOs reported by Braun et al. [78]. Fig. 3 suggests the possibilities to achieve high $E/k$ ratios in HECs/CCCs. Yet, we note that highest $E/k$ ratio reported to date is for $Dy_3NbO_7$ [123] (also labeled in Fig. 3), which is in the disordered fluorite structure (with random mixing of $Dy^{3+}$ and $Nb^{5+}$ in 3:1 ratios in one cation sublattice) with relative low mixing entropy of ~0.56$k_B$ but large size disorder of ~13.5% (defined in [14]; see further discussion in §4.1). This unique trait makes HECs attractive for applications such as thermal barrier coatings (TBCs) [124] and thermoelectrics [125].

In this section, we illustrate and discuss the new and unique opportunities brought by HECs and CCCs, as well as the potential benefits of broadening HECs to CCCs as we proposed, using thermally-insulating yet stiff CCCs as an example.

**4.1 Thermal Conductivity**

While most experimental HEC studies have reported lowered thermal conductivity, only a couple of reports have investigated this phenomenon in depth [13, 78]. Braun et al. used a virtual crystal approximation model to investigate the thermal conduction mechanism in the rocksalt-structured ESO, and they showed that ~50% reduction in thermal conductivity could be achieved when adding another cation in the ESO derivatives [78]. The authors ruled out mass and size disorder (by using nearby elements in the periodic table) and anharmonicity (via adopting components with similar thermal expansion coefficients as MgO). Consequently, they attributed the temperature-independent (amorphous or glass-like) thermal conductivity to disorder in the interatomic force constants (IFCs). This claim was further supported by an extended X-ray absorption fine structure analysis, which revealed a highly strained anion sublattice that presumably led to the suppressed thermal conductivity.

A couple of very recent research reports [13, 14] suggest broadening HECs to CCCs to achieve even lower thermal conductivity in MECs that can outperform their high-entropy counterparts.

On the one hand, the ideal mixing (configurational) entropy is not the best descriptor to describe thermal conductivity. For example, size disorder (see definition in Ref. [14]) has been proposed to be a more effective descriptor (than ideal mixing entropy itself) in describing thermal conductivity in medium- and high-entropy pyrochlore oxides (Fig. 4(a) and 4(b)) [14]. Theoretical work by Schelling et al. on simple cubic pyrochlore oxides has also suggested that the low conductivity is governed by size disorder [126]. The significant role of size disorder suggests an important role of severe lattice deformation in reducing thermal conductivity in HECs and CCCs. It also suggests that the lower thermal conductivities are not always coincident with high-entropy compositions, which was also confirmed experimentally in pyrochlore oxides, as shown in Fig. 4(a) [14]. The fact that $Dy_3NbO_7$, which has low ideal mixing entropy of ~0.56$k_B$ per cation but large size disorder of $\delta_{size} \approx 13.5\%$, exhibits the highest $E/k$ ratio reported to date [123] further supports this suggestion.

On the other hand, while the equimolar condition does provide the highest configurational entropy assuming ideal mixing, other variables such as oxygen vacancy concentration and cation valency may also be significant. This point is highlighted by Wright et al. in a study on non-equimolar fluorite oxides [13]. The authors found that the thermal conductivity of their YSZ-like fluorite oxides to be dependent on the nominal oxygen vacancy concentration $[V_o^{\cdot\cdot}]$, e.g., in a series of $(Hf_{1/3}Zr_{1/3}Ce_{1/3})_{1-x}(Y_{1/2}Yb_{1/2})_xO_{2-\delta}$ specimens shown in Fig. 5 [13]. More than approximately 5% of oxygen vacancies in the anion sublattice would likely lead to clustering and potentially ordering of the oxygen vacancies that suppress the point defect scattering, similar to that is well known for YSZ [127–129].

The combination of these studies [13, 14] demonstrates that various medium-entropy compositions (Fig. 1) can outperform their high-entropy counterparts, thereby supporting a step forward to broaden HECs to CCCs. Moreover, the increased compositional space in CCCs, which is significantly larger than that in HECs, allows for further engineering and design capabilities.

It is worth noting that Braun et al. [78] and Wright et al. [14] suggested different vital parameters controlling the thermal conductivity: disorders in charge/force constants vs. atomic/cation sizes. It is unlikely that one simple descriptor can be used to forecast the thermal conductivity in all HECs and CCCs. Furthermore, these two parameters are likely coupled. It should be noted that there has yet to be direct experiments to probe the role of the force constant variation in reducing thermal conductivity. This will need to be probed by spectrophotometry, vibrational spectroscopy, or electron spin resonance spectroscopy; however, the band overlap may occur for chemically similar elements, rendering such analysis difficult [130–133]. Further in-depth mechanistic studies are needed.

**4.2 Mechanical Properties**

Another interesting general observation of HECs is represented by their enhanced mechanical properties. The generally observed enhanced hardness from the RoM averages [44, 48, 50, 54, 56, 59, 61, 63, 64, 66] may be explained through solid-solution strengthening. Castle et al. also highlighted the importance of the activated slip systems affecting hardness and the relative easiness of activation and switching of dominant slip systems [61]. Additionally, the increase in chemical disorder may change the dominant slip, thereby resulting in variation in ductility and hardness [64]. The importance of mass and size disorder acting as an impedance and scattering the dislocation's group energy was suggested by Sarker et al. for high-entropy carbides [66].

Harrington et al. further noted that traditional solid-solution strengthening is likely significant; yet, the overall electronic band structure is equally (or potentially more) important [59]. The electronic structure arises from the bonding nature, and it has a significant impact on the available slip systems and direction in the material. Thus, the comparison of hardness between an HEC and the RoM of constituents is usually null. The improved mechanical properties may be due to solid-solution strengthening or perhaps unique available slip systems only accessible through increased configurational entropy. Models accounting for the solid-solution strengthening, electronic band structure, and Hall-Petch effects are warranted to explain observations further.

The hardness is typically directly related to Young's modulus ($E$). Interestingly, the modulus of HECs was also found to be enhanced in some cases [13, 20, 59, 61, 64, 66, 78]. The underlying mechanisms are unknown

In defective Compositionally Complex fluorite oxides, the moduli and hardness values are comparable with YSZ despite the addition of high fractions of soft stabilizers [13]. Fig. 5 shows that the modulus of $(Hf_{1/3}Zr_{1/3}Ce_{1/3})_{1-x}(Y_{1/2}Yb_{1/2})_xO_{2-\delta}$ specimens are roughly the same for equimolar compositions and 8YSZ-like $Hf_{0.284}Zr_{0.284}Ce_{0.284}Y_{0.074}Yb_{0.074}O_{2-\delta}$. However, the modulus drops substantially with further reduction in the amounts of stabilizers (presumably due to the instability of the cubic phase).

In high-entropy metal diborides, a recent study showed that incorporating softer $WB_2$ and $MoB_2$ components makes single-phase high-entropy borides harder, which suggested unusual and unexpected phenomena can occur in HECs [134].

**4.3 *E/k* Ratios**

Braun et al. used Young's modulus ($E$) to thermal conductivity ($k$) ratio, $E/k$, as a parameter to estimate the phonon scattering rate since $E/k \propto 1/C_v \tau$, where $C_v$ and $\tau$ are the volumetric heat capacity and phonon lifetime, respectively [78]. This parameter can serve as a figure of merit for the suppressed thermal conductivity and enhanced phonon scattering in HECs and CCCs. There exists a strong tradeoff between thermal conductivity and Young's modulus because both should increase with strong atomic bonding; however, HECs have been shown to have a unique capability to break this tradeoff with a record $E/k$ of 143.7 $GPa\,m\,K\,W^{-1}$ in 2018 [78]. Later, Yang et al. broke this record with a disordered fluorite oxide, Dy$_3$NbO$_7$ with relatively low entropy, $\sim 0.56 k_B$ per cation, but high size disorder of $\delta_{size} \approx 13.5\%$ (see the definition in Ref. [14]) that exhibits the highest $E/k$ of 235 $GPa\,m\,K\,W^{-1}$ [123] reported to date.

Notably, Wright et al. found that high $E/k$ ratios can be achieved in MECs (*i*) in non-equimolar defective fluorite oxides with an optimal amount of oxygen vacancies (Fig. 5) [13] and (*ii*) in ordered pyrochlore oxides with considerable size disorder (Fig. 4(c)) [14], both of which outperformed their high-entropy counterparts. High $E/k$ ratios achieved by several representative HECs and CCCs are shown in Fig. 3.

## 5. Concluding Remarks

Suppressed thermal conductivity and enhanced mechanical properties are likely inherent to HECs and CCCs. The $E/k$ ratios can serve as a useful figure of merit to guide the design of thermally-insulating, yet stiff ceramics for potential applications as new classes of TBCs (albeit that other properties such as the matching thermal expansion coefficients have also need to be considered). The classic trade-off in materials selections between the low thermal conductivity and high modulus and hardness can be broken via exploring HECs and CCCs, thereby suggesting exciting new opportunities. Further in-depth studies are needed to understand the underlying mechanisms and develop useful descriptors.

Until recently, nearly all the attention in HEC research has been placed on five-component (and, in a few cases, four-component) equimolar compositions. Here, we propose a step forward to expand HECs to CCCs (i.e., Compositionally Complex ceramics) to include medium-entropy and non-equimolar compositions. This proposal is inspired and supported by a couple of our most recent experimental studies showing that non-equimolar defective fluorite oxides with an optimal amount of oxygen vacancies (Fig. 5) [13] and medium-entropy pyrochlore oxides with considerable size disorder (Fig. 4(c)) [14] can both outperform their high-entropy counterparts.

Possible benefits that arise from the broadening of HECs to CCCs, as well as considering more complex CCCs with two or more cation sublattices and possibly anion-site mixing, are represented by the much increased compositional space (order of magnitude higher than the already vast compositional space of equimolar HECs) and the more degrees of freedom to tune properties, particularly multiple properties at the same time. Additional revenue of tailoring the defects and disorder is represented by introducing aliovalent cations and associated anion vacancies, e.g., in YSZ-like defective fluorite oxides (Fig. 2), and local order (CSROs), long-range disorder-order (e.g., fluorite-pyrochlore) transformation, and heterogeneity (e.g., nanodomains). The development of CCCs is also likely to reveal fascinating and improved properties in many other relevant areas such as catalysts, electrochemical performance, or corrosion resistance. In such a case, let us "embrace the complexity" to achieve superior and tunable properties.

In this proactive pursuit of "embracing the complexity," the diversifying classes of CCCs illustrated in Fig. 2 provide orders of magnitudes more possibilities than equimolar HECs to tailor the composition, defects, disorder, and short- and long-range order to achieve better and more tunable performance properties. However, the vast compositional space also poses a major challenge in designing HECs, which is even more challenging for CCCs. In this regard, in-depth studies of the underlying mechanisms and development of various descriptors and strategies to predict useful trends are warranted and essential.

We conclude by stating that a significant amount of progress in HEC research that has been obtained and the nearly exponential growth since 2015 are staggering. Ceramics have a niche use in the world, but their properties are duly unique, and their demands for improvements are ubiquitous. While HECs are still in their infancy, the broadened compositional space to CCCs will enable the continuing the research to be limitless. Again, let us "embrace the complexity!"

**6. Acknowledgment**

This material is primarily based upon work supported by the U.S. Department of Energy's Office of Energy Efficiency and Renewable Energy (EERE) under Solar Energy Technologies Office (SETO) Agreement Number EE0008529. This perspective article is also benefitted from the insights learned in other independent research partially supported by an ONR MURI program (N00014-15-1-2863), an EERE H2@Scale program (DE-EE0008839), and a Vannevar Bush Faculty Fellowship (N00014-16-1-2569) through cross-catalyzing. We thank Prof. Renkun Chen, Qingyang Wang, Mingde Qin, and Dr. Joshua Gild for their scientific insights via our collaborative research in the last few years via conducting several projects (acknowledged above) in various relevant areas.

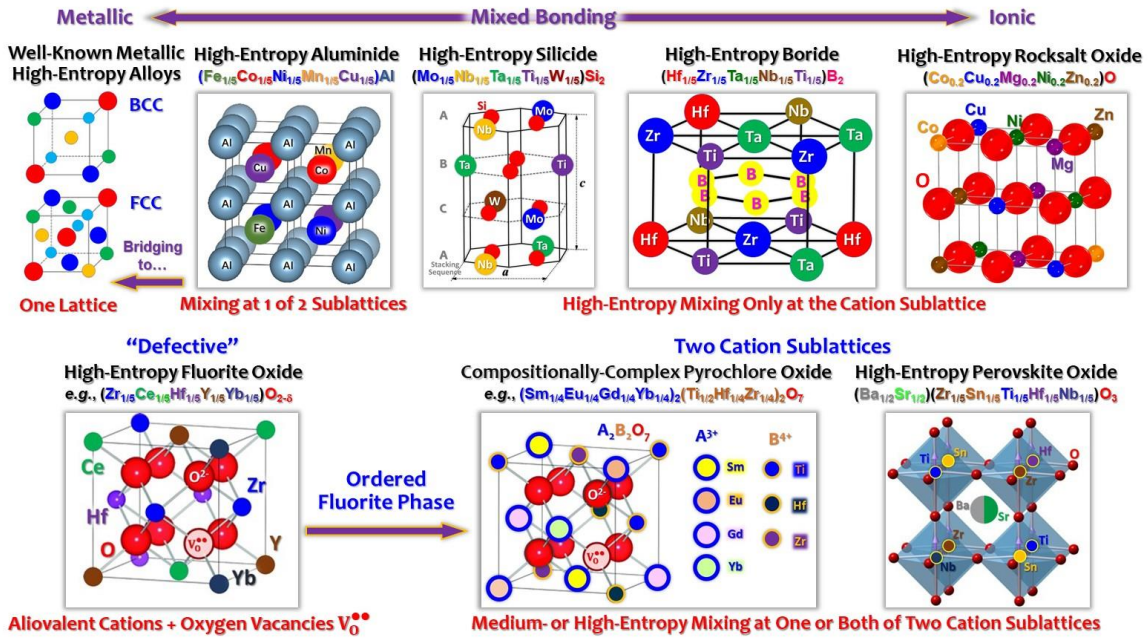

**Figure 1.** Schematic illustration of selected high-entropy ceramics (HECs) fabricated and reported in recent literature. From the well-known metallic high-entropy alloys (HEAs) to the newly reported high-entropy aluminides [43], silicides [44, 45], borides [46–56], carbides (not shown here) [21, 22, 57–66], and oxides [11, 13, 14, 23–37, 68–95]; the bonding character changes from metallic to mixed (metallic/covalent/ionic), and then to mostly ionic. Here, the discovery of single-phase, equimolar, high-entropy intermetallic compounds such as $(Fe_{1/5}Co_{1/5}Ni_{1/5}Mn_{1/5}Cu_{1/5})Al$ [43], which are structurally like HECs (with high-entropy mixing on one of two sublattices) but with mostly metallic (and some ionic) bonding, bridges the HEAs and HECs. Moreover, the fabrication of YSZ-like Compositionally Complex fluorite oxides with substantial amounts of aliovalent cations and oxygen vacancies, as well as ordered Compositionally Complex pyrochlore oxides and high-entropy perovskite oxides with two cationic sublattices, further broaden the diversifying families of HECs and CCCs.

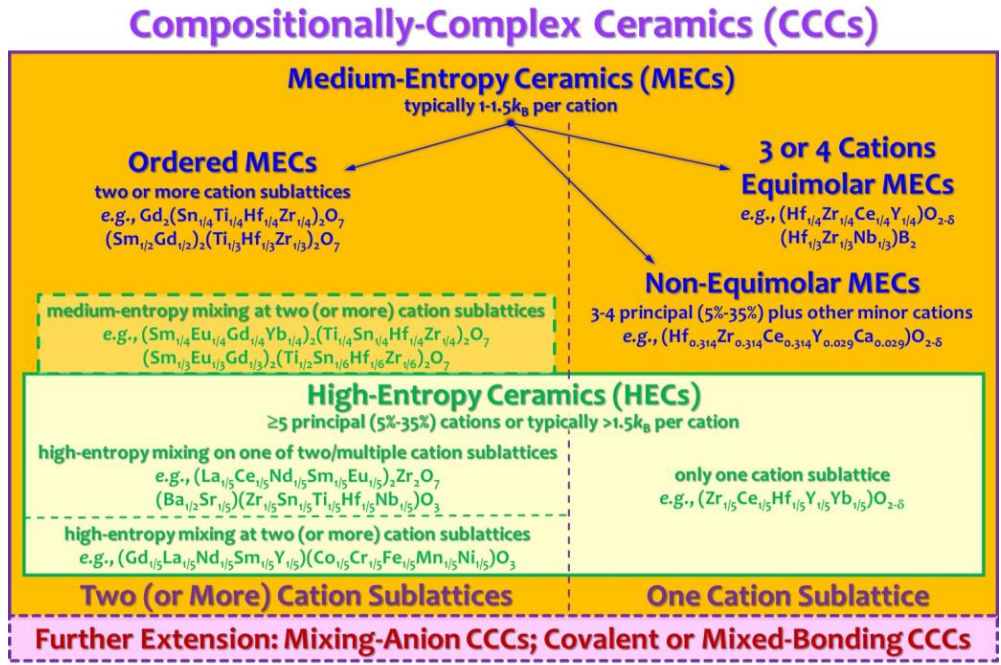

**Figure 2.** We propose to extend high-entropy ceramics (HECs) to Compositionally Complex ceramics (CCCs), also known as "multi-principal cation ceramics (MPCCs)" [13], to include medium-entropy ceramics (MECs) with typical mixing configurational entropy in the range of $1 - 1.5k_B$ per cation. Here, HECs are loosely defined as compositions of five or more principal (typically 5%–35%) cations in equimolar or near-equimolar fractions, with typically $> 1.5k_B$ per cation ideal mixing configurational entropy. MECs include (*i*) 3-4 cation, equimolar (or near-equimolar) compositions and (*ii*) non-equimolar compositions with 3-4 principal plus a few minor (typically <5%) cations. For crystal structures with two or more cation sublattices (e.g., perovskite and pyrochlore), HECs refer to compositions with high-entropy mixing in either one or more cation sublattice(s). Here, ceramic compositions with medium-entropy mixing on two (or more) cation sublattices may be loosely considered as an extension to HECs. Further extensions include mixing-anion CCCs and covalent (or mixed-bonding) CCCs. Finally, entropy-stabilized ceramics (ESCs) were defined separately [11], which can overlap with the definitions of CCCs, HECs, and MECs above.

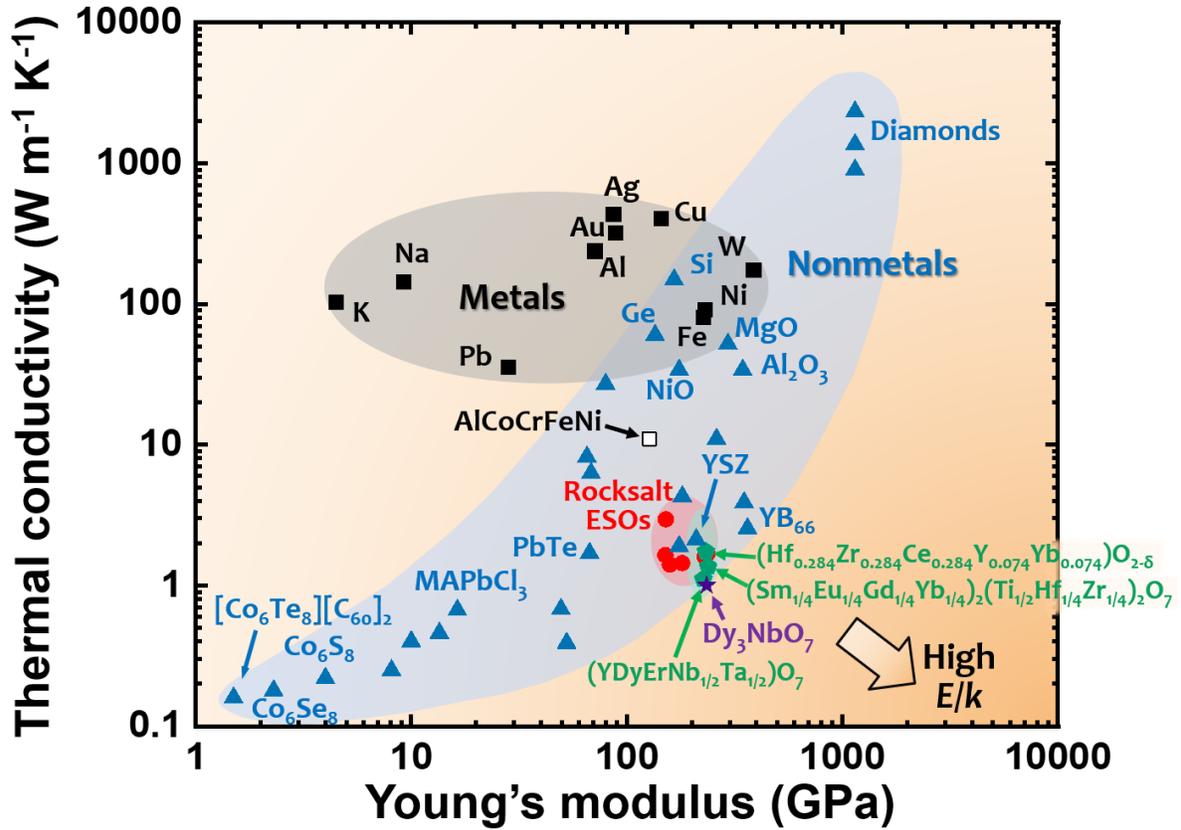

**Figure 3.** Thermal conductivity ($k$) vs. Young's modulus ($E$) in a double logarithmical plot to feature several HECs/CCCs with high $E/k$ ratio. The open square denotes a metallic HEA, AlCoCrFeNi, which exhibits a high $E/k$ ratio in comparison with pure metals. The black squares, blue triangles, and red discs denote metallic, nonmetallic, and rocksalt ESO systems, respectively. Replotted after Ref. [78] with permission (Copyright 2020, John Wiley and Sons), where we further added the data points for $Dy_3NbO_7$ (denoted by the purple star) [123] and three Compositionally Complex fluorite-based oxides (denoted by green pentagons): non-equimolar $Hf_{0.284}Zr_{0.284}Ce_{0.284}Y_{0.074}Yb_{0.074}O_{2-\delta}$ [13] and $(YDyErNb_{1/2}Ta_{1/2})O_7$ rare earth niobate/tantalate [our unpublished data] in the disordered fluorite structure, as well as $(Sm_{1/4}Eu_{1/4}Gd_{1/4}Yb_{1/4})_2(Ti_{1/2}Hf_{1/4}Zr_{1/4})_2O_7$ in the ordered pyrochlore structure [14].

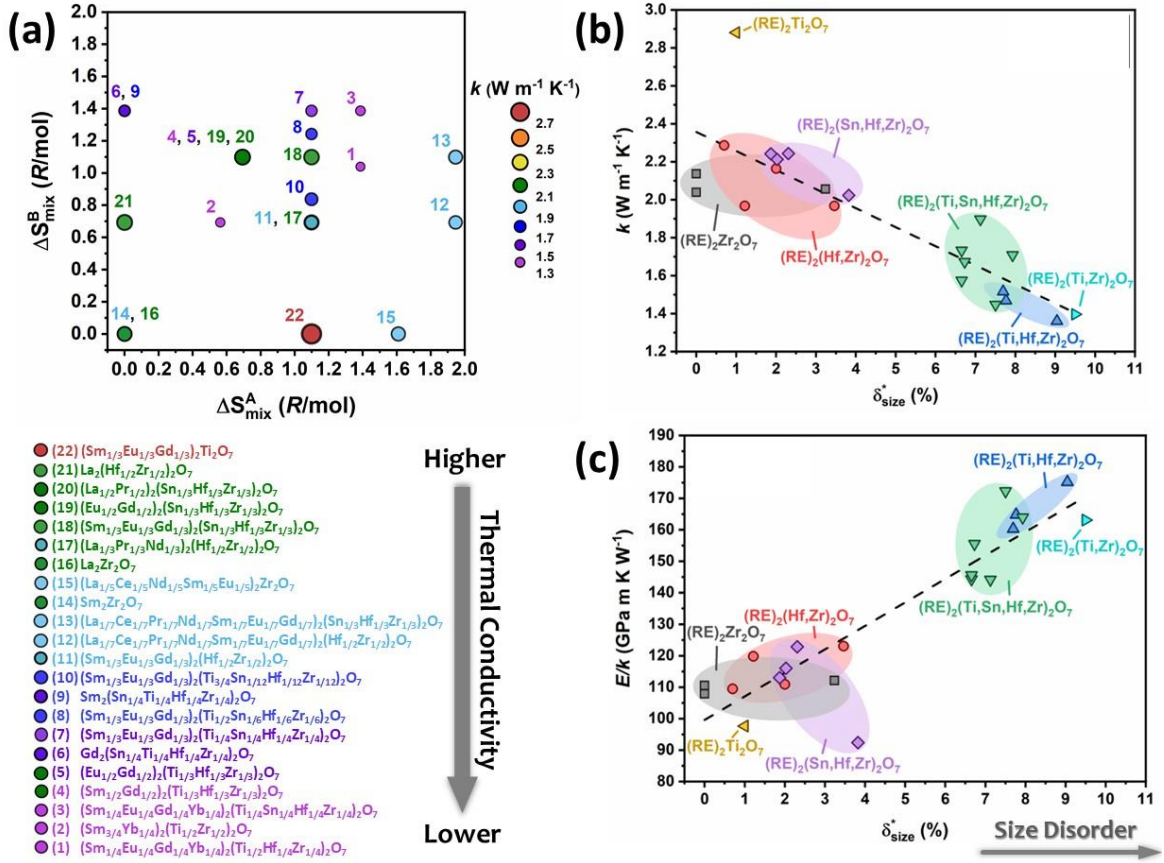

**Figure 4. (a)** The measured thermal conductivities (represented by color and size) of 22 single-phase pyrochlore oxides, plotted in the 2D space of ideal mixing configurational entropy in the A and B sublattices, respectively, using the data in Ref. [14]. Correlation of **(b)** thermal conductivity ($k$) and **(c)** the $E/k$ ratios of these 22 pyrochlores with the size disorder parameter, $\delta^*_{size}$. It was, therefore, suggested that the size disorder parameter $\delta^*_{size}$ (instead of the ideal mixing entropy itself) can be used as a more effective descriptor to forecast $k$ and the $E/k$ ratio in CCCs. Panel (b) and (c) reprinted from Ref. [14] with permission (Copyright 2020, Elsevier).

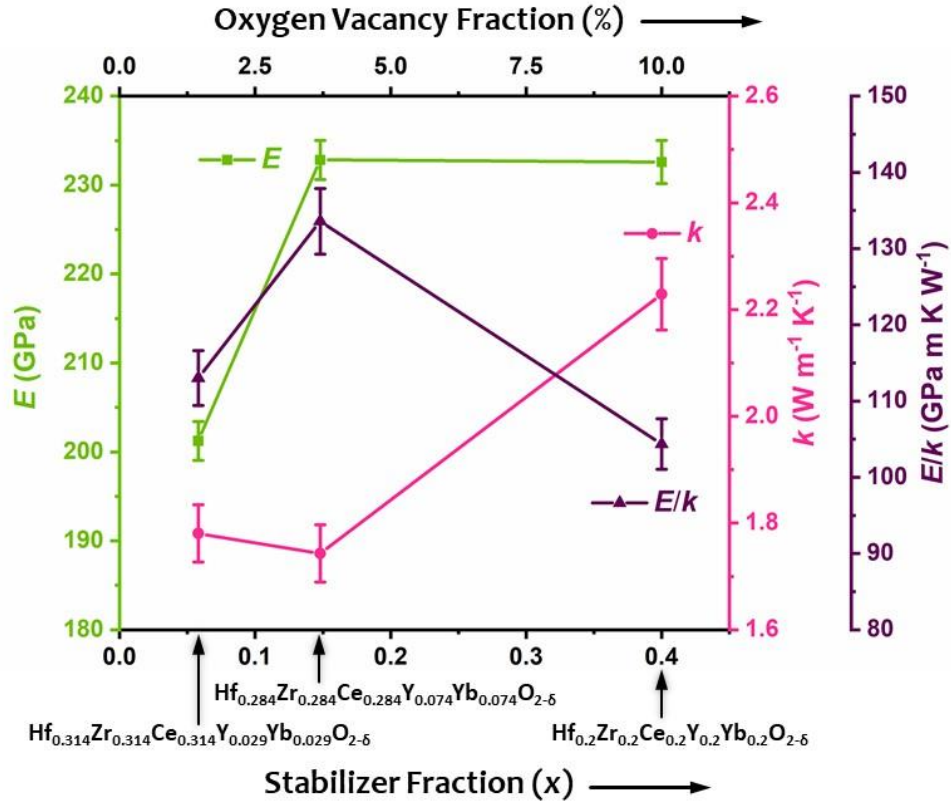

**Figure 5.** Measured Young's modulus ($E$), thermal conductivity ($k$), and $E/k$ ratio for a series of $(Hf_{1/3}Zr_{1/3}Ce_{1/3})_{1-x}(Y_{1/2}Yb_{1/2})_xO_{2-\delta}$ specimens, where $x$ is the total stabilizer fraction. Medium-entropy, non-equimolar specimen $Hf_{0.284}Zr_{0.284}Ce_{0.284}Y_{0.074}Yb_{0.074}O_{2-\delta}$ exhibits the lowest thermal conductivity and the highest $E/k$ ratio. Replotted with revisions from Ref. [13] with permission (Copyright 2020, Elsevier).

**References**

1. Zhang R-Z, Reece MJ (2019) Review of high entropy ceramics: design, synthesis, structure and properties. J Mater Chem A 7:22148–22162. https://doi.org/10.1039/C9TA05698J
2. Lei Z, Liu X, Wang H, et al (2019) Development of advanced materials via entropy engineering. Scr Mater 165:164–169. https://doi.org/10.1016/j.scriptamat.2019.02.015
3. Miracle DB, Senkov ON (2017) A critical review of high entropy alloys and related concepts. Acta Mater 122:448–511. https://doi.org/10.1016/j.actamat.2016.08.081
4. George EP, Raabe D, Ritchie RO (2019) High-entropy alloys. Nat Rev Mater 4:515–534. https://doi.org/10.1038/s41578-019-0121-4
5. Yeh JW, Chen SK, Lin SJ, et al (2004) Nanostructured high-entropy alloys with multiple principal elements: novel alloy design concepts and outcomes. Adv Eng Mater 6:299–303. https://doi.org/10.1002/adem.200300567
6. Cantor B, Chang ITH, Knight P, Vincent AJB (2004) Microstructural development in equiatomic multicomponent alloys. Mater Sci Eng A 375:213–218. https://doi.org/10.1016/j.msea.2003.10.257
7. Hsieh MH, Tsai MH, Shen WJ, Yeh JW (2013) Structure and properties of two Al-Cr-Nb-Si-Ti high-entropy nitride coatings. Surf Coatings Technol 221:118–123. https://doi.org/10.1016/j.surfcoat.2013.01.036
8. Yeh JW (2013) Alloy design strategies and future trends in high-entropy alloys. Jom 65:1759–1771. https://doi.org/10.1007/s11837-013-0761-6
9. Lai CH, Lin SJ, Yeh JW, Chang SY (2006) Preparation and characterization of AlCrTaTiZr multi-element nitride coatings. Surf Coatings Technol 201:3275–3280. https://doi.org/10.1016/j.surfcoat.2006.06.048
10. Yeh JW, Chen YL, Lin SJ, Chen SK (2007) High-entropy alloys - A new era of exploitation. Mater Sci Forum 560:1–9. https://doi.org/10.4028/www.scientific.net/MSF.560.1
11. Rost CM, Sachet E, Borman T, et al (2015) Entropy-stabilized oxides. Nat Commun 6:8485. https://doi.org/10.1038/ncomms9485
12. Oses C, Toher C, Curtarolo S (2020) High-entropy ceramics. Nat Rev Mater 1–15. https://doi.org/10.1038/s41578-019-0170-8

13. Wright AJ, Wang Q, Huang C, et al (2020) From high-entropy ceramics to Compositionally Complex ceramics: A case study of fluorite oxides. J Eur Ceram Soc. https://doi.org/10.1016/j.jeurceramsoc.2020.01.015
14. Wright AJ, Wang Q, Ko S-T, et al (2020) Size disorder as a descriptor for predicting reduced thermal conductivity in medium- and high-entropy pyrochlores. Scr Mater. https://doi.org/10.1016/j.scriptamat.2020.02.011
15. Gali A, George EP (2013) Tensile properties of high- and medium-entropy alloys. Intermetallics 39:74–78. https://doi.org/10.1016/j.intermet.2013.03.018
16. Gludovatz B, Hohenwarter A, Thurston KVS, et al (2016) Exceptional damage-tolerance of a medium-entropy alloy CrCoNi at cryogenic temperatures. Nat Commun 7:10602. https://doi.org/10.1038/ncomms10602
17. Li Z, Raabe D (2017) Strong and ductile non-equiatomic high-entropy alloys: Design, processing, microstructure, and mechanical properties. Jom 69:2099–2106. https://doi.org/10.1007/s11837-017-2540-2
18. Zhao Z, Xiang H, Dai FZ, et al (2019) (TiZrHf)P2O7: An equimolar multicomponent or high entropy ceramic with good thermal stability and low thermal conductivity. J. Mater. Sci. Technol. 35:2227–2231
19. Heng C, Huimin X, Fu-Zhi D, et al (2019) High entropy (Yb0.25Y0.25Lu0.25Er0.25)2SiO5 with strong anisotropy in thermal expansion. J Mater Sci Technol. https://doi.org/10.1016/j.jmst.2019.07.022
20. Ren X, Tian Z, Zhang J, Wang J (2019) Equiatomic quaternary (Y1/4Ho1/4Er1/4Yb1/4)2SiO5 silicate: A perspective multifunctional thermal and environmental barrier coating material. Scr Mater 168:47–50. https://doi.org/10.1016/j.scriptamat.2019.04.018
21. Dusza J, Švec P, Girman V, et al (2018) Microstructure of (Hf-Ta-Zr-Nb)C high-entropy carbide at micro and nano/atomic level. J Eur Ceram Soc 38:4303–4307. https://doi.org/10.1016/j.jeurceramsoc.2018.05.006
22. Demirskyi D, Borodianska H, Suzuki TS, et al (2019) High-temperature flexural strength performance of ternary high-entropy carbide consolidated via spark plasma sintering of TaC, ZrC and NbC. Scr Mater 164:12–16. https://doi.org/10.1016/j.scriptamat.2019.01.024

23. Gild J, Samiee M, Braun JL, et al (2018) High-entropy fluorite oxides. J Eur Ceram Soc 38:3578–3584. https://doi.org/10.1016/j.jeurceramsoc.2018.04.010

24. Sharma Y, Musico BL, Gao X, et al (2018) Single-crystal high entropy perovskite oxide epitaxial films. Phys Rev Mater 2:. https://doi.org/10.1103/PhysRevMaterials.2.060404

25. Sarkar A, Djenadic R, Wang D, et al (2018) Rare earth and transition metal based entropy stabilised perovskite type oxides. J Eur Ceram Soc 38:2318–2327. https://doi.org/10.1016/j.jeurceramsoc.2017.12.058

26. Jiang S, Hu T, Gild J, et al (2018) A new class of high-entropy perovskite oxides. Scr Mater 142:116–120. https://doi.org/10.1016/j.scriptamat.2017.08.040

27. Biesuz M, Fu S, Dong J, et al (2019) High entropy Sr((Zr0.94Y0.06)0.2Sn0.2Ti0.2Hf0.2Mn0.2)O3−x perovskite synthesis by reactive spark plasma sintering. J Asian Ceram Soc 7:127–132. https://doi.org/10.1080/21870764.2019.1595931

28. Witte R, Sarkar A, Kruk R, et al (2019) High-entropy oxides: An emerging prospect for magnetic rare-earth transition metal perovskites. Phys Rev Mater 3:034406. https://doi.org/10.1103/PhysRevMaterials.3.034406

29. Dąbrowa J, Stygar M, Mikuła A, et al (2018) Synthesis and microstructure of the (Co,Cr,Fe,Mn,Ni)3O4 high entropy oxide characterized by spinel structure. Mater Lett 216:32–36. https://doi.org/10.1016/j.matlet.2017.12.148

30. Grzesik Z, Smoła G, Miszczak M, et al (2020) Defect structure and transport properties of (Co,Cr,Fe,Mn,Ni)3O4 spinel-structured high entropy oxide. J Eur Ceram Soc 40:835–839. https://doi.org/10.1016/j.snb.2019.127065

31. Mao A, Quan F, Xiang HZ, et al (2019) Facile synthesis and ferrimagnetic property of spinel (CoCrFeMnNi)3O4 high-entropy oxide nanocrystalline powder. J Mol Struct 1194:11–18. https://doi.org/10.1016/j.molstruc.2019.05.073

32. Mao A, Xiang H, Zhang Z, et al (2020) A new class of spinel high-entropy oxides with controllable magnetic properties. J Magn Magn Mater 165884. https://doi.org/10.1016/j.jmmm.2019.165884

33. Ren K, Wang Q, Shao G, et al (2020) Multicomponent high-entropy zirconates with comprehensive properties for advanced thermal barrier coating. Scr Mater 178:382–386. https://doi.org/10.1016/j.scriptamat.2019.12.006

34. Zhao Z, Xiang H, Dai F-Z, et al (2019) (La0.2Ce0.2Nd0.2Sm0.2Eu0.2)2Zr2O7: A novel high-entropy ceramic with low thermal conductivity and sluggish grain growth rate. J Mater Sci Technol. https://doi.org/10.1016/j.jmst.2019.05.054
35. Zhang K, Li W, Zeng J, et al (2020) Preparation of (La0.2Nd0.2Sm0.2Gd0.2Yb0.2)2Zr2O7 high-entropy transparent ceramic using combustion synthesized nanopowder. J Alloys Compd 817:153328. https://doi.org/10.1016/j.jallcom.2019.153328
36. Li F, Zhou L, Liu J-X, et al (2019) High-entropy pyrochlores with low thermal conductivity for thermal barrier coating materials. J Adv Ceram 1–7. https://doi.org/10.1007/s40145-019-0342-4
37. Teng Z, Zhu L, Tan Y, et al (2019) Synthesis and structures of high-entropy pyrochlore oxides. J Eur Ceram Soc. https://doi.org/10.1016/j.eplepsyres.2019.106192
38. Wen T, Liu H, Ye B, et al (2019) High-entropy alumino-silicides: a novel class of high-entropy ceramics. Sci China Mater 1–7. https://doi.org/10.1007/s40843-019-9585-3
39. Li F, Bao W, Sun SK, et al (2020) Synthesis of single-phase metal oxycarbonitride ceramics. Scr Mater 176:17–22. https://doi.org/10.1016/j.scriptamat.2019.09.024
40. Liu R, Chen H, Zhao K, et al (2017) Entropy as a gene-like performance indicator promoting thermoelectric materials. Adv Mater 29:1702712. https://doi.org/10.1002/adma.201702712
41. Fan Z, Wang H, Wu Y, et al (2017) Thermoelectric performance of PbSnTeSe high-entropy alloys. Mater Res Lett 5:187–194. https://doi.org/10.1080/21663831.2016.1244116
42. Fan Z, Wang H, Wu Y, et al (2016) Thermoelectric high-entropy alloys with low lattice thermal conductivity. RSC Adv 6:52164–52170. https://doi.org/10.1039/c5ra28088e
43. Zhou N, Jiang S, Huang T, et al (2019) Single-phase high-entropy intermetallic compounds (HEICs): bridging high-entropy alloys and ceramics. Sci Bull 64:856–864. https://doi.org/10.1016/j.scib.2019.05.007
44. Gild J, Braun J, Kaufmann K, et al (2019) A high-entropy silicide: (Mo0.2Nb0.2Ta0.2Ti0.2W0.2)Si2. J Mater. https://doi.org/10.1016/j.jmat.2019.03.002
45. Qin Y, Liu JX, Li F, et al (2019) A high entropy silicide by reactive spark plasma sintering. J Adv Ceram 8:148–152. https://doi.org/10.1007/s40145-019-0319-3

46. Mayrhofer PH, Kirnbauer A, Ertelthaler P, Koller CM (2018) High-entropy ceramic thin films; A case study on transition metal diborides. Scr Mater 149:93–97. https://doi.org/10.1016/j.scriptamat.2018.02.008

47. Liu D, Liu H, Ning S, et al (2019) Synthesis of high-purity high-entropy metal diboride powders by boro/carbothermal reduction. J Am Ceram Soc 102:7071–7076. https://doi.org/10.1111/jace.16746

48. Gild J, Zhang Y, Harrington T, et al (2016) High-entropy metal diborides: A new class of high-entropy materials and a new type of ultrahigh temperature ceramics. Sci Rep 6:37946. https://doi.org/10.1038/srep37946

49. Liu D, Wen T, Ye B, Chu Y (2019) Synthesis of superfine high-entropy metal diboride powders. Scr Mater 167:110–114. https://doi.org/10.1016/j.scriptamat.2019.03.038

50. Zhang Y, Jiang Z Bin, Sun SK, et al (2019) Microstructure and mechanical properties of high-entropy borides derived from boro/carbothermal reduction. J Eur Ceram Soc 39:3920–3924. https://doi.org/10.1016/j.jeurceramsoc.2019.05.017

51. Tallarita G, Licheri R, Garroni S, et al (2019) High-entropy transition metal diborides by reactive and non-reactive spark plasma sintering: A comparative investigation. J Eur Ceram Soc

52. Gild J, Kaufmann K, Vecchio K, Luo J (2019) Reactive flash spark plasma sintering of high-entropy ultrahigh temperature ceramics. Scr Mater 170:106–110. https://doi.org/10.1016/j.scriptamat.2019.05.039

53. Shen XQ, Liu JX, Li F, Zhang GJ (2019) Preparation and characterization of diboride-based high entropy (Ti0.2Zr0.2Hf0.2Nb0.2Ta0.2)B2–SiC particulate composites. Ceram Int 45:24508–24514. https://doi.org/10.1016/j.ceramint.2019.08.178

54. Gild J, Wright A, Quiambao-tomko K, et al (2019) Thermal conductivity and hardness of three single-phase high-entropy metal diborides fabricated by borocarbothermal reduction and spark plasma sintering. Ceram Int. https://doi.org/10.1016/j.ceramint.2019.11.186

55. Tallarita G, Licheri R, Garroni S, et al (2019) Novel processing route for the fabrication of bulk high-entropy metal diborides. Scr Mater 158:100–104. https://doi.org/10.1016/j.scriptamat.2018.08.039

56. Zhang Y, Guo WM, Jiang Z Bin, et al (2019) Dense high-entropy boride ceramics with ultra-high hardness. Scr Mater 164:135–139.

https://doi.org/10.1016/j.scriptamat.2019.01.021

57. Zhou J, Zhang J, Zhang F, et al (2018) High-entropy carbide: A novel class of multicomponent ceramics. Ceram Int 44:22014–22018. https://doi.org/10.1016/j.ceramint.2018.08.100

58. Wei X, Qin Y, Liu J, et al (2019) Gradient microstructure development and grain growth inhibition in high- entropy carbide ceramics prepared by reactive spark plasma sintering. J Eur Ceram Soc. https://doi.org/10.1016/j.jeurceramsoc.2019.12.034

59. Harrington TJ, Gild J, Sarker P, et al (2019) Phase stability and mechanical properties of novel high entropy transition metal carbides. Acta Mater 166:271–280. https://doi.org/10.1016/j.actamat.2018.12.054

60. Wang YP, Gan GY, Wang W, et al (2018) Ab initio prediction of mechanical and electronic properties of ultrahigh temperature high-entropy ceramics (Hf0.2Zr0.2Ta0.2M0.2Ti0.2)B2 (M = Nb, Mo, Cr). Phys Status Solidi Basic Res 255:1800011. https://doi.org/10.1002/pssb.201800011

61. Castle E, Csanádi T, Grasso S, et al (2018) Processing and properties of high-entropy ultra-high temperature carbides. Sci Rep 8:8609. https://doi.org/10.1038/s41598-018-26827-1

62. Han X, Girman V, Sedlák R, et al (2019) Improved creep resistance of high entropy transition metal carbides. J Eur Ceram Soc

63. Zhang H, Hedman D, Feng P, et al (2019) A high-entropy B4(HfMo2TaTi)C and SiC ceramic composite. Dalt Trans 48:5161–5167. https://doi.org/10.1039/c8dt04555k

64. Csanádi T, Castle E, Reece MJ, Dusza J (2019) Strength enhancement and slip behaviour of high-entropy carbide grains during micro-compression. Sci Rep 9:1–14. https://doi.org/10.1038/s41598-019-46614-w

65. Yan X, Constantin L, Lu Y, et al (2018) (Hf0.2Zr0.2Ta0.2Nb0.2Ti0.2)C high-entropy ceramics with low thermal conductivity. J Am Ceram Soc 101:4486–4491. https://doi.org/10.1111/jace.15779

66. Sarker P, Harrington T, Toher C, et al (2018) High-entropy high-hardness metal carbides discovered by entropy descriptors. Nat Commun 9:1–10. https://doi.org/10.1038/s41467-018-07160-7

67. Jin T, Sang X, Unocic RR, et al (2018) Mechanochemical-assisted synthesis of high-

entropy metal nitride via a soft urea strategy. Adv Mater 30:1–5. https://doi.org/10.1002/adma.201707512

68. Moballegh A, Rost CM, Maria J-P, Dickey EC (2015) Chemical homogeneity in entropy-stabilized complex metal oxides. Microsc Microanal 21:1349–1350. https://doi.org/10.1017/S1431927615007539

69. Bérardan D, Franger S, Meena AK, Dragoe N (2016) Room temperature lithium superionic conductivity in high entropy oxides. J Mater Chem A 4:9536–9541. https://doi.org/10.1039/c6ta03249d

70. Bérardan D, Franger S, Dragoe D, et al (2016) Colossal dielectric constant in high entropy oxides. Phys Status Solidi - Rapid Res Lett 10:328–333. https://doi.org/10.1002/pssr.201600043

71. Meisenheimer PB, Kratofil TJ, Heron JT (2017) Giant enhancement of exchange coupling in entropy-stabilized oxide heterostructures. Sci Rep 7:13344. https://doi.org/10.1038/s41598-017-13810-5

72. Bérardan D, Meena AK, Franger S, et al (2017) Controlled Jahn-Teller distortion in (MgCoNiCuZn)O-based high entropy oxides. J Alloys Compd 704:693–700. https://doi.org/10.1016/j.jallcom.2017.02.070

73. Rost CM, Rak Z, Brenner DW, Maria JP (2017) Local structure of the $Mg_xNi_xCo_xCu_xZn_xO(x=0.2)$ entropy-stabilized oxide: An EXAFS study. J Am Ceram Soc 100:2732–2738. https://doi.org/10.1111/jace.14756

74. Sarkar A, Djenadic R, Usharani NJ, et al (2017) Nanocrystalline multicomponent entropy stabilised transition metal oxides. J Eur Ceram Soc 37:747–754. https://doi.org/10.1016/j.jeurceramsoc.2016.09.018

75. Sarkar A, Loho C, Velasco L, et al (2017) Multicomponent equiatomic rare earth oxides with a narrow band gap and associated praseodymium multivalency. Dalt Trans 46:12167–12176. https://doi.org/10.1039/c7dt02077e

76. Djenadic R, Sarkar A, Clemens O, et al (2017) Multicomponent equiatomic rare earth oxides. Mater Res Lett 5:102–109. https://doi.org/10.1080/21663831.2016.1220433

77. Kotsonis GN, Rost CM, Harris DT, Maria J-P (2018) Epitaxial entropy-stabilized oxides: growth of chemically diverse phases via kinetic bombardment. MRS Commun 8:1371–1377. https://doi.org/10.1557/mrc.2018.184

78. Braun JL, Rost CM, Lim M, et al (2018) Charge-induced disorder controls the thermal conductivity of entropy-stabilized oxides. Adv Mater 30:1805004. https://doi.org/10.1002/adma.201805004
79. Chen H, Fu J, Zhang P, et al (2018) Entropy-stabilized metal oxide solid solutions as CO oxidation catalysts with high-temperature stability. J Mater Chem A 6:11129–11133. https://doi.org/10.1039/c8ta01772g
80. Anand G, Wynn AP, Handley CM, Freeman CL (2018) Phase stability and distortion in high-entropy oxides. Acta Mater 146:119–125. https://doi.org/10.1016/j.actamat.2017.12.037
81. Biesuz M, Spiridigliozzi L, Dell'Agli G, et al (2018) Synthesis and sintering of (Mg, Co, Ni, Cu, Zn)O entropy-stabilized oxides obtained by wet chemical methods. J Mater Sci 53:8074–8085. https://doi.org/10.1007/s10853-018-2168-9
82. Zhai S, Rojas J, Ahlborg N, et al (2018) The use of poly-cation oxides to lower the temperature of two-step thermochemical water splitting. Energy Environ Sci 11:2172–2178. https://doi.org/10.1039/c8ee00050f
83. Sarkar A, Velasco L, Wang D, et al (2018) High entropy oxides for reversible energy storage. Nat Commun 9:3400. https://doi.org/10.1038/s41467-018-05774-5
84. Chen K, Pei X, Tang L, et al (2018) A five-component entropy-stabilized fluorite oxide. J Eur Ceram Soc 38:4161–4164. https://doi.org/10.1016/j.jeurceramsoc.2018.04.063
85. Jimenez-Segura MP, Takayama T, Bérardan D, et al (2019) Long-range magnetic ordering in rocksalt-type high-entropy oxides. Appl Phys Lett 114:122401. https://doi.org/10.1063/1.5091787
86. Hong W, Chen F, Shen Q, et al (2019) Microstructural evolution and mechanical properties of (Mg,Co,Ni,Cu,Zn)O high-entropy ceramics. J Am Ceram Soc 102:2228–2237. https://doi.org/10.1111/jace.16075
87. Vinnik DA, Trofimov EA, Zhivulin VE, et al (2019) High-entropy oxide phases with magnetoplumbite structure. Ceram Int 45:12942–12948. https://doi.org/10.1016/j.ceramint.2019.03.221
88. Mao A, Xiang HZ, Zhang ZG, et al (2019) Solution combustion synthesis and magnetic property of rock-salt ($Co_{0.2}Cu_{0.2}Mg_{0.2}Ni_{0.2}Zn_{0.2}$)O high-entropy oxide nanocrystalline powder. J Magn Magn Mater 484:245–252. https://doi.org/10.1016/j.jmmm.2019.04.023

89. Zhang J, Yan J, Calder S, et al (2019) Long-range antiferromagnetic order in a rocksalt high entropy oxide. Chem Mater 31:3705–3711. https://doi.org/10.1021/acs.chemmater.9b00624

90. Qiu N, Chen H, Yang Z, et al (2019) A high entropy oxide (Mg0.2Co0.2Ni0.2Cu0.2Zn0.2O) with superior lithium storage performance. J Alloys Compd 777:767–774. https://doi.org/10.1016/j.jallcom.2018.11.049

91. Dupuy AD, Wang X, Schoenung JM (2019) Entropic phase transformation in nanocrystalline high entropy oxides. Mater Res Lett 7:60–67. https://doi.org/10.1080/21663831.2018.1554605

92. Osenciat N, Bérardan D, Dragoe D, et al (2019) Charge compensation mechanisms in Li-substituted high-entropy oxides and influence on Li superionic conductivity. J Am Ceram Soc 102:6156–6162. https://doi.org/10.1111/jace.16511

93. Wang Q, Sarkar A, Li Z, et al (2019) High entropy oxides as anode material for Li-ion battery applications: A practical approach. Electrochem commun 100:121–125. https://doi.org/10.1016/j.elecom.2019.02.001

94. Chellali MR, Sarkar A, Nandam SH, et al (2019) On the homogeneity of high entropy oxides: An investigation at the atomic scale. Scr Mater 166:58–63. https://doi.org/10.1016/j.scriptamat.2019.02.039

95. Zheng Y, Yi Y, Fan M, et al (2019) A high-entropy metal oxide as chemical anchor of polysulfide for lithium-sulfur batteries. Energy Storage Mater. https://doi.org/10.1016/j.ensm.2019.02.030

96. Parida T, Karati A, Guruvidyathri K, et al (2020) Novel rare-earth and transition metal-based entropy stabilized oxides with spinel structure. Scr Mater 178:513–517. https://doi.org/10.1016/j.scriptamat.2019.12.027

97. Qin M, Gild J, Hu C, et al (2020) High-entropy boride-carbide two-phase ultrahigh temperature ceramics fabricated by reactive spark plasma sintering. arXiv prepirint arXiv: 2002.09756, submitted to J. Eur. Ceram. Soc. https://arxiv.org/abs/2002.09756

98. Zhang R-Z, Gucci F, Zhu H, et al (2018) Data-driven design of ecofriendly thermoelectric high-entropy sulfides. Inorg Chem 57:13027–13033. https://doi.org/10.1021/acs.inorgchem.8b02379

99. Chen X, Wu Y (2019) High-entropy transparent fluoride laser ceramics. J Am Ceram Soc

103:750–756. https://doi.org/10.1111/jace.16842

100. Chen H, Zhao Z, Xiang H, et al (2020) Effect of reaction routes on the porosity and permeability of porous high entropy (Y0.2Yb0.2Sm0.2Nd0.2Eu0.2)B6 for transpiration cooling. J Mater Sci Technol 38:80–85. https://doi.org/10.1016/j.jmst.2019.09.006

101. de la Obra AG, Avilés MA, Torres Y, et al (2017) A new family of cermets: Chemically complex but microstructurally simple. Int J Refract Met Hard Mater 63:17–25. https://doi.org/10.1016/j.ijrmhm.2016.04.011

102. Vinnik DA, Trofimov EA, Zhivulin VE, et al (2019) The new extremely substituted high entropy (Ba,Sr,Ca,La)Fe6-x(Al,Ti,Cr,Ga,In,Cu,W)xO19 microcrystals with magnetoplumbite structure. Ceram Int. https://doi.org/10.1016/j.ceramint.2019.12.232

103. Xu W, Chen H, Jie K, et al (2019) Entropy-driven mechanochemical synthesis of polymetallic zeolitic imidazolate frameworks for CO2 fixation. Angew Chemie 131:5072–5076. https://doi.org/10.1002/ange.201900787

104. Radoń A, Hawełek Ł, Łukowiec D, et al (2019) Dielectric and electromagnetic interference shielding properties of high entropy (Zn,Fe,Ni,Mg,Cd)Fe2O4 ferrite. Sci Rep 9:1–13. https://doi.org/10.1038/s41598-019-56586-6

105. Zhao Z, Chen H, Xiang H, et al (2019) (La0.2Ce0.2Nd0.2Sm0.2Eu0.2)PO4: A high-entropy rare-earth phosphate monazite ceramic with low thermal conductivity and good compatibility with Al2O3. J Mater Sci Technol 35:2892–2896. https://doi.org/10.1016/j.jmst.2019.08.012

106. Dong Y, Ren K, Lu Y, et al (2019) High-entropy environmental barrier coating for the ceramic matrix composites. J Eur Ceram Soc 39:2574–2579. https://doi.org/10.1016/j.jeurceramsoc.2019.02.022

107. Lei Z, Liu X, Li R, et al (2018) Ultrastable metal oxide nanotube arrays achieved by entropy-stabilization engineering. Scr Mater 146:340–343. https://doi.org/10.1016/j.scriptamat.2017.12.025

108. Giri A, Braun JL, Rost CM, Hopkins PE (2017) On the minimum limit to thermal conductivity of multi-atom component crystalline solid solutions based on impurity mass scattering. Scr Mater 138:134–138. https://doi.org/10.1016/j.scriptamat.2017.05.045

109. Giri A, Braun JL, Hopkins PE (2018) Reduced dependence of thermal conductivity on temperature and pressure of multi-atom component crystalline solid solutions. J Appl Phys

123:015106. https://doi.org/10.1063/1.5010337

110. Lim M, Rak Z, Braun JL, et al (2019) Influence of mass and charge disorder on the phonon thermal conductivity of entropy stabilized oxides determined by molecular dynamics simulations. J Appl Phys 125:055105. https://doi.org/10.1063/1.5080419

111. Rák Z, Maria JP, Brenner DW (2018) Evidence for Jahn-Teller compression in the (Mg, Co, Ni, Cu, Zn)O entropy-stabilized oxide: A DFT study. Mater Lett 217:300–303. https://doi.org/10.1016/j.matlet.2018.01.111

112. Yang Y, Wang W, Gan GY, et al (2018) Structural, mechanical and electronic properties of (TaNbHfTiZr)C high entropy carbide under pressure: Ab initio investigation. Phys B Condens Matter 550:163–170. https://doi.org/10.1016/j.physb.2018.09.014

113. Ye B, Wen T, Huang K, et al (2019) First-principles study, fabrication, and characterization of ($Hf_{0.2}Zr_{0.2}Ta_{0.2}Nb_{0.2}Ti_{0.2}$)C high-entropy ceramic. J Am Ceram Soc 102:4344–4352. https://doi.org/10.1111/jace.16295

114. Ye B, Wen T, Nguyen MC, et al (2019) First-principles study, fabrication and characterization of ($Zr_{0.25}Nb_{0.25}Ti_{0.25}V_{0.25}$)C high-entropy ceramics. Acta Mater 170:15–23. https://doi.org/10.1016/j.actamat.2019.03.021

115. Dai F-Z, Wen B, Sun Y, et al (2020) Theoretical prediction on thermal and mechanical properties of high entropy ($Zr_{0.2}Hf_{0.2}Ti_{0.2}Nb_{0.2}Ta_{0.2}$)C by deep learning potential. J Mater Sci Technol

116. Sarkar A, Wang Q, Schiele A, et al (2019) High-entropy oxides: Fundamental aspects and electrochemical properties. Adv Mater 31:1806236. https://doi.org/10.1002/adma.201806236

117. Cheng B, Lou H, Sarkar A, et al (2019) Pressure-induced tuning of lattice distortion in a high-entropy oxide. Commun Chem 2:1–9. https://doi.org/10.1038/s42004-019-0216-2

118. Padture NP (2016) Advanced structural ceramics in aerospace propulsion. Nat Mater 15:804–809. https://doi.org/10.1038/nmat4687

119. Binner J, Porter M, Baker B, et al (2019) Selection, processing, properties and applications of ultra-high temperature ceramic matrix composites, UHTCMCs – a review. Int Mater Rev 1–56. https://doi.org/10.1080/09506608.2019.1652006

120. Ye B, Wen T, Chu Y (2020) High-temperature oxidation behavior of ($Hf_{0.2}Zr_{0.2}Ta_{0.2}Nb_{0.2}Ti_{0.2}$)C high-entropy ceramics in air. J Am Ceram Soc 103:500–

507. https://doi.org/10.1111/jace.16725

121. Ye B, Wen T, Liu D, Chu Y (2019) Oxidation behavior of (Hf0.2Zr0.2Ta0.2Nb0.2Ti0.2)C high-entropy ceramics at 1073-1473 K in air. Corros Sci 153:327–332. https://doi.org/10.1016/j.corsci.2019.04.001

122. Chen H, Xiang H, Dai FZ, et al (2019) High porosity and low thermal conductivity high entropy (Zr0.2Hf0.2Ti0.2Nb0.2Ta0.2)C. J Mater Sci Technol 35:1700–1705. https://doi.org/10.1016/j.jmst.2019.04.006

123. Yang J, Qian X, Pan W, et al (2019) Diffused Lattice Vibration and Ultralow Thermal Conductivity in the Binary Ln–Nb–O Oxide System. Adv Mater 31:1808222. https://doi.org/10.1002/adma.201808222

124. Padture NP, Gell M, Jordan EH (2002) Thermal barrier coatings for gas-turbine engine applications. Science (80- ) 296:280–284. https://doi.org/10.1126/science.1068609

125. Snyder GJ, Toberer ES (2010) Complex thermoelectric materials. In: Materials for Sustainable Energy: A Collection of Peer-Reviewed Research and Review Articles from Nature Publishing Group. World Scientific Publishing Co., pp 101–110

126. Schelling PK, Phillpot SR, Grimes RW (2004) Optimum pyrochlore compositions for low thermal conductivity. Philos Mag Lett 84:127–137. https://doi.org/10.1080/09500830310001646699

127. Wachsman ED (2004) Effect of oxygen sublattice order on conductivity in highly defective fluorite oxides. J Eur Ceram Soc 24:1281–1285. https://doi.org/10.1016/S0955-2219(03)00509-0

128. Steele D, Fender EF (1974) The structure of cubic ZrO2:YO1.5 solid solutions by neutron scattering. J Phys C Solid State Phys 7:1

129. Bisson J-F, Fournier D, Poulain M, et al (2004) Thermal conductivity of yttria-zirconia single crystals, determined with spatially resolved infrared thermography. J Am Ceram Soc 83:1993–1998. https://doi.org/10.1111/j.1151-2916.2000.tb01502.x

130. Suárez CB, Grinfeld R (1970) Contribution to the Study of GdO Band Spectrum. J Chem Phys 53:1110–1117. https://doi.org/10.1063/1.1674106

131. Huber KP, Herzberg G (1979) Constants of Diatomic Molecules. Van Nostrand Reinhold Company, New York

132. Hess NJ, Begg BD, Conradson SD, et al (2002) Spectroscopic investigations of the

structural phase transition in Gd2(Ti1-yZry)2O7 pyrochlores. J Phys Chem B 106:4663–4677. https://doi.org/10.1021/jp014285t

133. Li CW, Hong J, May AF, et al (2015) Orbitally driven giant phonon anharmonicity in SnSe. Nat Phys 11:1063–1069. https://doi.org/10.1038/nphys3492

134. Qin M, Gild J, Wang H, et al (2019) Stabilizing softer WB2 and MoB2 components into high-entropy borides via boron-metals reactive sintering to attain higher hardness. arXiv prepirint arXiv:1912.11743, submitted to J. Eur. Ceram. Soc. https://arxiv.org/abs/1912.11743